# Emergence of superconductivity in strongly correlated hole-dominated $Fe_{1-x}Se$

S. L. Ni,[1,2] J. P. Sun,[1,2] S. B. Liu,[1,2] J. Yuan,[1,6] Li Yu,[1,2,7] M. W. Ma,[1] L. Zhang,[5] L. Pi,[5] P. Zheng,[1] P. P. Shen,[1,2] D. Li,[1,2] D. E. Shi,[4] G. B. Li,[4*] J. L. Sun,[4] G. M. Zhang,[3] K. Jin,[1,2,6,7] J.-G. Cheng,[1,2,7] F. Zhou,[1,2*] X. L. Dong,[1,2,6,7*] and Z. X. Zhao[1,2,6,7]

[1] *Beijing National Laboratory for Condensed Matter Physics and Institute of Physics, Chinese Academy of Sciences, Beijing 100190, China*

[2] *School of Physical Sciences, University of Chinese Academy of Sciences, Beijing 100049, China*

[3] *State Key Laboratory of Low Dimensional Quantum Physics and Department of Physics, Tsinghua University, Beijing 100084, China*

[4] *College of Chemistry and Molecular Engineering, Peking University, Beijing National Laboratory for Molecular Sciences (BNLMS), State Key Laboratory of Rare Earth Materials Chemistry and Applications, Beijing, 100871, China*

[5] *Anhui Key Laboratory of Condensed Matter Physics at Extreme Conditions, High Magnetic Field Laboratory, Chinese Academy of Sciences, Hefei 230031, China*

[6] *Key Laboratory for Vacuum Physics, University of Chinese Academy of Sciences, Beijing 100049, China*

[7] *Songshan Lake Materials Laboratory, Dongguan, Guangdong 523808, China*

**Abstract**

Here we establish a more complete phase diagram for FeSe system, based on experimental results of nonstoichiometric $Fe_{1-x}Se$ single crystals that we have developed recently, as well as nearly stoichiometric FeSe single crystals. The electronic correlation is found to be strongly enhanced in hole-dominated $Fe_{1-x}Se$, as compared with electron-dominated FeSe, from the magnetic susceptibility and electrical transport measurements in the normal state. A superconducting dome is found to emerge starting from the strongly correlated hole-dominated regime with electron doping, while the tetragonal-orthorhombic phase transition at ~90 K is observed only at higher electron-doping levels in the electron-dominated regime.

Among iron-based superconductors, the simplest binary FeSe has drawn continuous research attention due to its unique properties in the normal and superconducting (SC) state [1,2]. Signs of the electronic correlation effects have been observed experimentally in bulk FeSe [1]. Model calculation [3] has predicted that reducing the *d*-electron number per iron ($N_d$) by introducing holes can further enhance the electronic correlations of Hund's coupling [4] and Coulomb interaction, as well as iron magnetic moment, as it drives the iron-based superconductors (approximate $3d^6$ configuration) closer to the Mott insulators (at half-filled $3d^5$ configuration). Due to limited sample availability, however, previous investigations of FeSe system are mainly based on the nearly undoped FeSe samples near $N_d = 6$, which show superconducting transition at an almost constatnt $T_c \sim 9$ K and strongly electron-dominated transport regime in the normal state. Therefore, an experimental study of the doping effects on the electronic correlation in FeSe system is still lacking to date. A way out of this is to introduce holes into the FeSe system by creating vacant Fe-sites. Correspondingly, the *d*-electron number in resultant nonstoichiometric $Fe_{1-x}Se$ samples is reduced to $N_d = 6 - p$ depending on their iron deficiency $x$; here $p = 2x/(1-x)$ is the doped hole concentration relative to that ($p = 0$) of stoichiometric FeSe where $x = 0$. This simplest $Fe_{1-x}Se$ series can provide a superior platform for experimentally studying the doping dependence of electronic correlation and its influence on superconductivity. On the other hand, bulk FeSe does not order magnetically, and has been shown to display significant anisotropic antiferromagnetic (AFM) fluctuations [5-8]. It undergoes a high-temperature tetragonal (HTT) to low-temperature orthorhombic (LTO) structural transition upon cooling to $T_s \sim 80 - 90$ K [9,10]. The orthorhombic distortion is weak, as charaterized by a small amount of in-plane lattice change $(a-b)/(a+b) \sim 0.3\%$, and has been proposed as driven by the development of a long-range nematic order with unequal occupations of iron $3d_{xz}$ and $3d_{yz}$ orbitals [11-13]. This is dissimilar to related iron arsenide systems with a HTT to LTO phase transition closely followed by a long-range AFM ordering [14,15]. These different leading instabilities in iron selenide and arsenide compounds complicate a unified understanding of iron-based superconductivity. Nematic order in nearly stoichiometric FeSe has been intensively studied experimentally [1,16-20] and theoretically [21-23] in connection with superconductivity; however, considerable controversy still remains. Therefore, it is also necessary to investigate this important issue in the iron-deficient $Fe_{1-x}Se$ samples.

Here, we present the results of our experimental study on a series of nonstoichiometric $Fe_{1-x}Se$ and nearly stoichiometric FeSe single crystals. All the iron-deficient $Fe_{1-x}Se$ samples of varying $T_c$ exhibit a hole (h) dominated transport behavior at low temperatures in the normal state, in contrast to the electron (e) dominated behavior [24,25] in the nearly stoichiometric FeSe. Our measurements show that the electronic correlation and magnetic susceptibility are strongly enhanced in the h-dominated $Fe_{1-x}Se$ having a fractional number of reduction in the $d$-electron number $N_d$, in agreement with the previous calculation results. Importantly, the superconductivity appears deeply inside the strongly correlated h-dominated regime, and $T_c$ rises gradually from zero to 9 K with weakening strength of electronic correlation, which is associated with increasing electron doping (*i.e.* increasing $N_d$), or correspondingly decreasing concentration $p$ of the introduced holes. As $p \to 0$ with further electron doping, FeSe system crosses from the h-dominated $Fe_{1-x}Se$ to e-dominated FeSe, then $T_c$ tends to be reduced from the optimal value of 9 K. The results reveal a superconducting dome in the FeSe system, which emerges starting from the strongly correlated hole-dominated region, in analogy with that in high-$T_c$ cuprate superconductors derived from the strongly correlated Mott insulators. Interestingly, the onset doping level (at an Fe-deficiency $x = x_{sc} \sim 5.3\ \%$) for superconductivity to emerge in the prototypical $Fe_{1-x}Se$ is comparable to that (at $x_{sc} \sim 5\ \%$ [26]) in the intercalated high-$T_c$ (Li,Fe)OH$Fe_{1-x}$Se superconductors. On the other hand, the HTT-LTO phase transition is observed only at higher electron-doping levels in the e-dominated regime of nearly stoichiometric FeSe. This suggests that the electronic states developing with the long-range orbital-nematic order do not favor the superconductivity. Overall, the present study makes an important step forward toward an understanding of unconventional iron-based superconductivity in connection with the electronic correlation.

Phase-pure and sizable iron-deficient $Fe_{1-x}Se$ single crystals were synthesized by a hydrothermal ion-deintercalation (HID) method as reported previously (see Supplemental Material [27]). For comparison, nearly stoichiometric single- and poly-crystalline FeSe samples were also obtained by chemical vapor transport (CVT) and solid-state reaction methods [27], respectively. $T_c$ of all the samples was characterized by the onset temperature of sharp SC transition in magnetic susceptibility, as shown in the upper panel of Fig. 1a. The lower panel of Fig. 1a shows the SC transition in the resistivity. Here H9, H7.2, *etc.* stand for the HID samples with $T_c$ = 9 K, 7.2 K, *etc.*, respectively, and C8.6 stands for the CVT sample with $T_c$ = 8.6 K. By single-crystal x-ray structural refinements (see supplemental Table S1), the concentrations of disordered Fe-vacancies are determined for representative samples as $x \sim 2.5\ \%$ for H7.2, $\sim 1.8\ \%$ for H9, and $\sim 0.6\ \%$ for C8.6. CVT sample C8.6 is closer to the stoichiometry.

It is unexpected that the HTT-LTO phase transition is not observed in the HID $Fe_{1-x}Se$ single crystals, as clearly evidenced by the disappearance of specific-heat jump (Fig. 2a). In contrast, the specific-heat jump is evident and strong in the CVT (C8.6) and polycrystalline (S8, $T_c$ = 8 K) FeSe, indicating the structural transition at $T_s$ in these nearly stoichiometric samples. Further crystallographic evidence also suggests the absence of orthorhombic phase in the HID samples. With a single tetragonal (220) peak down to 20 K (Fig. 2d), the low-temperature powder x-ray diffraction (XRD) patterns of the HID samples (see Fig. 2c for H7.2) can be well refined with the tetragonal structure by Rietveld method (see Table S2), with no acceptable orthorhombic solution obtained. This is in consistence with the specific-heat results. By comparison, the splitting of HTT (220) peak into LTO doublet (400) and (040) below $T_s$ and stronger reduction in the peak heights at 20 K (Fig. 2e) than the HID $Fe_{1-x}Se$ H7.2 (Fig. 2d) are also characteristic of the orthorhombic phase in the polycrystalline sample S8. For CVT FeSe sample C8.6, a kink at the structural transition $T_s \sim 90$ K is clearly visible in the resistivity $\rho(T)$ curve (Fig. 1b), similar to previous results; however, this kink is hardly discernible in the HID $Fe_{1-x}Se$ single crystals. The relative broadening of XRD peak profiles for H7.2 (Fig. 2d), as compared with the HTT phase of S8 (Fig. 2e), mainly results from the effects of lattice strain and grain size based on Williamson-Hall analysis [27]. But, the grain-size effect is present only in the ground powders (for the powder XRD experiments) but not the as-grown single crystals of HID $Fe_{1-x}Se$. The atomic ratio Fe/Se of HID samples obtained by the single-crystal structural refinements (Table S1) is smaller than that of nearly stoichiometric samples, including the present samples C8.6, S8 and the samples of previous work [5,28-31] as well. The results are summarized in Fig. 2b by plotting Fe/Se ratio against $T_c$. Thus, a boundary is identified between the samples having (Fe/Se $\gtrsim$ 0.99) or not having (Fe/Se $\lesssim$ 0.99) the orthorhombic phase transition.

Distinctive electrical transport behavior is also observed in the HID $Fe_{1-x}Se$ series at low temperatures in the normal state. The Hall resistivity $\rho_{xy}(H)$ of the HID single crystals is almost linearly proportional to magnetic field $H$ at all the measuring temperatures; the data of the representative H7.2 are given in Supplemental Material [27]. Similar linear $\rho_{xy}(H)$ was observed in FeSe films [32], however the film structural properties were not easy to characterize. For nearly stoichiometric CVT FeSe sample C8.6, a concave nonlinearity in $\rho_{xy}(H)$ develops below ~70 K [27], which is consistent with previous results. The corresponding temperature dependences of Hall coefficient $R_H$ are summarized in Fig. 3a. Previous work has shown a nearly compensated electrical transport in CVT FeSe [1,2]. Here as evident from Fig. 3a, above $T \sim 70$ K, all the HID $Fe_{1-x}Se$ samples share almost the same temperature dependence of $R_H$ with the CVT FeSe samples, though more holes are introduced into HID $Fe_{1-x}Se$ than CVT FeSe (with about $10^{20}$ cm$^{-3}$ holes doped by 1 % Fe deficiency $x$). The HID and CVT $R_H(T)$ curves are all close to zero and well coincide with each other, with the sign changed twice at $T = 120$ and 200 K. This means the nearly compensated electrical transport in both the HID and CVT samples, and the sign changes of their $R_H(T)$ are caused by an imbalance between the hole and electron mobilities. In fact, the strongly negative $R_H(T)$ of CVT FeSe at low temperatures has been ascribed to an emergence of a small minority of highly mobile electron carriers [24,25]. For the HID $Fe_{1-x}Se$ samples, however, their Hall coefficient $R_H(T)$ at low $T$ (*e.g.* 10 K) are significantly positive, and their values of $R_H(10 K)$ show a strong doping dependence (Fig. 4, the middle panel). This suggests the predominating hole-mobility contribution at low temperatures, especially in the higher-$T_c$ $Fe_{1-x}Se$ samples showing the larger positive $R_H(10 K)$ values, where the hole-channel is much more conductive than the electron-channel, despite their lower hole-concentration $p$ (smaller Fe-deficiency $x$) as shown in Fig. 4. The compensated feature of the HID $Fe_{1-x}Se$ samples is also consistent with their almost linear $\rho_{xy}(H)$ at all the meauring temperatures [27] and their nonmonotonic variation in $R_H(T)$ (Fig. 3a). Here we note that recent results of upper critical field $H_{c2}(T)$ have also revealed a two-band feature of this HID $Fe_{1-x}Se$ series [33]. Interestingly, due to the stronger hole dominanc in the higher-$T_c$ ($\gtrsim$ 7.2 K) $Fe_{1-x}Se$ single crystals, their $H_{c2}(T)$ can also be fitted by a single-band Werthamer-Helfand-Hohenberg (WHH) formula, in addition to the effective two-band description. In the lower panel of Fig. 4, we present all the Hall coefficient data by a contour plot for the HID and CVT samples. This plot gives a whole perspective of the temperature and doping dependences of the carrier characteristics in the normal state of FeSe system. At higher temperatures above ~70 K, all the HID and CVT samples share the common feature in Hall signal, regardless of their different electron-doping level. At lower temperatures closer to $T_c$, however, distinctly different regions of the strong hole and electron dominance develop in the HID and CVT samples, respectively; and they are divided from each other at an Fe-deficiency $x = x_{h/e}$ (~1.4 %), corresponding to an electron-doping level at $N_d = N_d^{h/e}$ (~5.97).

In addition to the very different structural and transport properties, unusual magnetic properties in the normal state are further revealed in the HID single crystals. As presented in Fig. 3b for the representative samples, the value of magnetic susceptibility $\chi(T)$ for hole-dominated $Fe_{1-x}Se$ H7.2 is one order of magnitude larger than that for electron-dominated FeSe C8.6. And the temperature dependence displays a pronounced maximum centered around 70 K, below which it decreases linearly on cooling and approximately follows the Curie-Weiss behavior far beyond. Moreover, a consistently large effective local moment in the magnitude of ~3 $\mu_B$ is estimated for $Fe_{1-x}Se$ H7.2 from the Curie-Weiss law, $\chi_{cw}= C/(T+\Theta)$. A large iron magnetic moment is expected for the enhanced electronic correlation [3,4] associated with the reduced $d$-electron number $N_d$ in the hole-dominated $Fe_{1-x}Se$. Similarly, such enhanced correlation was also observed in heavily hole-doped iron arsenide superconductor $KFe_2As_2$ (at $x = 1$ in $Ba_{1-x}K_xFe_2As_2$ series) [34]. In that case a local magnetic moment of about 2.5 $\mu_B$ was also inferred from the Curie-Weiss law.

Moreover, the prominent crossover in $\chi(T)$ from the linear-$T$ to Curie-Weiss behavior around 70 K (Fig. 3b) is a manifestation of the enhanced correlation in the hole-dominated $Fe_{1-x}Se$ as well. Similar crossover feature was also observed in strongly correlated hole-doped iron arsenide $AFe_2As_2$ (A = K, Rb, Cs) superconductors, where the crossover occurs around a temperature of 100 K [34,35]. But the crossover feature sharply contrasts with the monotonic linear decrease in $\chi(T)$ all the way from room temperature down to low temperatures seen in electron-dominated FeSe C8.6 (Fig. 3b), which is consistent with previous results of CVT FeSe [36], and has also been observed in some other iron arsenide superconductors [37]. The linear reduction in magnetic susceptibility with lowering temperature is understood from the short-range AFM correlations of local magnetic moments [37]. When the thermal excitation can overcome such short-range magnetic correlations, the magnetic susceptibility $\chi(T)$ will turn from the linear-$T$ to Curie-Weiss behavior. For those iron-based superconductors with weaker electronic correlation, the strength of magnetic coupling is expected to be stronger, thus pushing the crossover feature to a higher temperature [37] beyond the ordinary temperature.

Furthermore, the doping dependence of electronic correlation is also reflected in the transport measurements at low temperatures just above $T_c$. In the hole-dominated regime, first note that an upturn in resistivity $\rho(T)$ develops below ~15 K (see Fig. 1b) in HID single crystals H3 and H4.6 with lower electron doping (*i.e.* smaller *d*-electron number $N_d$; see the upper panel of Fig. 4). This is a typical signature of carrier localization, as commonly seen in strongly correlated systems. Moreover, the resistivity at low $T = 10$ K displays a large doping-depedent increase among the HID samples, by a factor of 20 from H9 to H3. Simultaneously, the value of Hall coefficient at 10 K, $R_H(10$ K$)$, also changes significantly with doping (Fig. 4, the middle panel). Such changes in the low-$T$ Hall coefficient and resistivity cannot be accounted for by the Fe-vacancy scattering alone. First, the variation in concentration $x$ of the Fe-vacancies is relatively very small (just a few per cent); and second, the Fe-vacancies have equal scattering effects on the electron and hole carriers, therefore they themselves can not cause such strong doping-dependent change in $R_H(10$ K$)$ of the nearly compensated $Fe_{1-x}Se$ series. Further, due to the stronger hole dominance in the higher-$T_c$ ($\gtrsim$ 7.2 K) HID $Fe_{1-x}Se$ single crystals, their upper critical field $H_{c2}(T)$ can also be fitted by the WHH formula as mentioned above, which yields an enhancement of the effective mass $m^*$ with decreasing $N_d$ [33]. Therefore, the strong doping dependences of the low-$T$ Hall coefficient $R_H(10$ K$)$ and resistivity $\rho(10$ K$)$ can be attributed mainly to the electronic correlation effects, which are enhanced more and more in the hole-dominated regime. Thus a strong renormalization of low-$T$ quasiparticles are expected, associated with the enhancements of the effective mass $m^*$ and scattering rate $1/\tau$ of such quasiparticles due to the enhanced electronic correlation. The significant decrease in $R_H(10$ K$)$ and large increase in $\rho(10\ K)$ can be explained as a result of the strong reduction in the carrier mobility $\mu$ ($\propto \tau/m^*$) with decreasing $d$-electron number $N_d$.

In the lower panel of Fig. 4, we plot the phase diagram of FeSe system by temperature $T$ vs. lattice constant $c$, across the iron-deficient Fe$_{1-x}$Se and nearly stoichiometric FeSe. The $c$-axis cell parameter can be used as a measure of the degree of Fe-deficiency $x$, since introducing Fe-vacancies reduces the lattice constant $c$ (see supplemental Fig. S5), while the change in the in-plane lattice parameters is much smaller. With increasing electron doping or increasing $N_d$ (correspondingly decreasing $p$ and increasing $c$), FeSe system shifts from the h-dominated ($N_d = 6 - p < 6$ in HID Fe$_{1-x}$Se) toward e-dominated (near $N_d = 6$ in CVT FeSe as $p \to 0$) regime. Correspondingly, the Hall coefficient at low $T = 10$ K just above $T_c$ increases with $N_d$ in the h-dominated regime. The superconductivity appears near $N_d = N_d^{sc} \sim 5.89$ (at an Fe-deficiency $x = x_{sc} \sim 5.3\ \%$), deeply inside the h-dominated regime where the electronic correlation is strongly enhanced. And then $T_c$ rises steadily with weakening strength of the correlation. As the system crosses from the h- into e-dominated regime in the close vicinity of $N_d^{h/e} \sim 5.97$ (at $x_{h/e} \sim 1.4\ \%$), $T_c$ reaches the optimal value of 9 K smoothly. The onset Fe-deficiency ($x_{sc} \sim 5.3\ \%$) for superconductivity in the prototypical Fe$_{1-x}$Se is comparable to that ($x_{sc} \sim 5\ \%$ [26]) in the intercalated (Li,Fe)OHFe$_{1-x}$Se of a high $T_c$ up to 42 K. This implies a universal doping level for superconductivity to emerge within the conducting FeSe-layers of the related iron-based compounds.

Upon entering the electron-dominated regime, however, $T_c$ of FeSe system becomes saturated at 9 K initially then prone to a subtle decrease, despite the further weakened electronic correlation with increasing $N_d$. Simultaneously, the orthorhombic phase transition at $T_s$ is observed only in this regime with $N_d \gtrsim N_d^{h/e}$ (or $x \lesssim x_{h/e}$), as summarized in Fig. 4. And highly mobile electron-carriers appear at low temperatures [24,25], responsible for the sudden sign change of low-$T$ Hall coefficient near $N_d^{h/e}$ ($x_{h/e}$) and strongly negative $R_H(10\ K)$ values (Fig. 4, the middle panel). Therefore, the saturation and subsequent decrease in $T_c$ seem to suggest that new electronic states developing with the long-range orbital-nematic order in the e-dominated regime do not favor the superconductivity stemming from the h-dominated regime, while the existing electron pairing channel is reserved. In addition, the high pressure $P$, as a clean control parameter to suppress the nematicity without changing the chemical stoichiometry, can further increase $T_c$'s (at the zero resistivity) of both the HID Fe$_{1-x}$Se and CVT FeSe to higher values, and they are all maximized under the same optimum pressure $P_{opti.} \sim 6$ GPa (see Fig. 3c). This is in consistence with a similar superconducting pairing mechanism for the HID and CVT samples as speculated above.

In high pressures, a new magnetic/orthorhombic phase is induced below $T_m$ in the CVT samples [38,39], which, however, is hardly discernible in pressurized HID samples [27]. As can be seen from Fig. 3c, the magnetic/orthorhombic phase seems also to slow down the development of superconductivity in CVT FeSe in the pressure range from 1.7 to 5 GPa, as compared with that in HID Fe$_{1-x}$Se H9 of the same $T_c$ (9 K at ambient pressure). Once this magnetic phase in the CVT sample starts to be suppressed at $P \gtrsim 5$ GPa, its $T_c$ exhibits a steep increase to an optimal value of 38.3 K around $P_{opti.}$, where the Hall coefficient at low $T$ becomes strongly positive [40]. In contrast, the pressure dependences of $T_c$ for HID samples evolve much more smoothly than the CVT sample, which reach the optimal $T_c$ values (26 K for H9 and 21 K for H7.2) under the same $P_{opti.}$. The orbital (below $T_s$) and pressure-induced magnetic (below $T_m$) long-range orders do not cooperate with the superconductivity. Therefore, the cleaner region where the long-range orders are suppressed could be more promising in probing into the key ingredient for superconducting pairing.

The phase diagram in Fig. 4 reveals a doping-dependent superconducting dome in FeSe system, which emerges starting from the strongly correlated region. This is analogous to that observed in cuprate superconductors [41,42], which are referred to as doped Mott insulators. Nevertheless, there are significant differences between them due to their different $3d$ electronic configurations. Firstly, in cuprates, the superconductivity emerges upon electron or hole doping in close proximity to AFM half-filled ($3d^9$ of $Cu^{2+}$) insulating parent state; and $T_c$ of electron-doped (Nd/Pr)$_{2-x}$Ce$_x$CuO$_{4-y}$ appears at a finite initial value [41,42]. By comparison, for nearly stoichiometric FeSe (near $3d^6$ of $Fe^{2+}$), proximity to the half-filled Mott limit (at $3d^5$ of $Fe^{3+}$) requires a high degree of Fe deficiency in Fe$_{1-x}$Se, which is $x \to 1/3$ (33 %), i.e. $p \to 1$, for $3d^6 \to 3d^5$. Actually, AFM localized behavior [43] has been observed in nanosheets of Fe$_4$Se$_5$ (with $x$ = 20 %, and average $N_d$ = 5.5). And ferrimagnetic (because of unequal moments of sublattices with AFM alignments) metallic state [44] has been reported for bulk Fe$_7$Se$_8$ ($x$ = 12.5 %, and $N_d$ ~ 5.71), which is relatively closer to the superconducting dome (starting at $x_{sc}$ ~ 5.3 %, and $N_d^{sc}$ ~ 5.89). Both the non-SC phase near the Fe-deficiency $x = x_{sc}$ and SC phase at higher electron-doping levels than the present samples deserve further studies. Secondly, there exists another important difference. Unlike the cuprates, the iron-based superconductors are multi-orbital/band systems because of the $3d^6$ configuration of five Fe-$3d$ orbitals and small crystal field splitting [45]. Therefore Hund's rule coupling becomes relevant in the iron-based materials [4]. This gives rise to the multiplicity and complexity in iron-based superconductors, such as orbital selectivity and spin-orbit coupling.

To summarize, we establish a more complete joint phase diagram for FeSe system,

across the hole-dominated iron-deficient $Fe_{1-x}Se$ and electron-dominated nearly stoichiometric FeSe. The superconductivity emerges deeply inside the hole-dominated regime where the electronic correlation is strongly enhanced, and $T_c$ evolves in a dome-like shape with increasing electron doping and weakening strength of electronic correlation. The onset doping level for superconductivity to emerge seems to be universal for FeSe-based superconductors. The tetragonal-orthorhombic phase transition is observed only at higher electron-doping levels in the electron-dominated regime, where $T_c$ tends to decrease with further electron doping. The experimental findings presented here have important implication for the underlying physics of iron-based superconductivity in connection with the electronic correlation. Further work is required for a microscopic understanding of FeSe system in analogy with cuprate superconductors.

**Acknowledgements**
We would like to thank Prof. Jiangping Hu for helpful discussion, and Drs. L. H. Yang and J. Li for technical assistance. This work was supported by National Natural Science Foundation of China (Nos.11888101, 11834016), the National Key Research and Development Program of China (Grant Nos. 2017YFA0303003, 2016YFA0300300, 2016YFA0301004), and the Strategic Priority Research Program and Key Research Program of Frontier Sciences of the Chinese Academy of Sciences (Grant Nos. QYZDY-SSW-SLH013, QYZDY-SSW-SLH001).

\* Correspondence to: fzhou@iphy.ac.cn (F.Z.); liguobao@pku.edu.cn (G.B.L.); dong@iphy.ac.cn (X.L.D.)

**References**

[1] A. I. Coldea and M. D. Watson, Annu. Rev. Condens. Matter Phys. **9**, 125 (2018).

[2] D.-H. Lee, Annu. Rev. Condens. Matter Phys. **9**, 261 (2018).

[3] T. Misawa, K. Nakamura, and M. Imada, Phys. Rev. Lett. **108**, 177007 (2012).

[4] Z. P. Yin, K. Haule, and G. Kotliar, Nat. Mater. **10**, 932 (2011).

[5] Q. S. Wang, Y. Shen, B. Y. Pan, Y. Q. Hao, M. W. Ma, F. Zhou, P. Steffens, K. Schmalzl, T. R. Forrest, M. Abdel-Hafiez, X. J. Chen, D. A. Chareev, A. N. Vasiliev, P. Bourges, Y. Sidis, H. B. Cao, and J. Zhao, Nat. Mater. **15**, 159 (2016).

[6] M. C. Rahn, R. A. Ewings, S. J. Sedlmaier, S. J. Clarke, and A. T. Boothroyd, Phys. Rev. B **91**, 180501 (R) (2015).

[7] Q. S. Wang, Y. Shen, B. Y. Pan, X. W. Zhang, K. Ikeuchi, K. Iida, A. D. Christianson, H. C. Walker, D. T. Adroja, M. Abdel-Hafiez, X. J. Chen, D. A. Chareev, A. N. Vasiliev, and J. Zhao, Nat. Commun. **7**, 12182 (2016).

[8] T. Chen, Y. Chen, A. Kreisel, X. Lu, A. Schneidewind, Y. Qiu, J. T. Park, T. G. Perring, J. R. Stewart, H. Cao, R. Zhang, Y. Li, Y. Rong, Y. Wei, B. M. Andersen, P. J. Hirschfeld, C. Broholm, and P. Dai, Nat. Mater. **18**, 709 (2019).

[9] S. Margadonna, Y. Takabayashi, M. T. McDonald, K. Kasperkiewicz, Y. Mizuguchi, Y. Takano, A. N. Fitch, E. Suard, and K. Prassides, Chem. Commun. **0**, 5607 (2008).

[10] T. M. McQueen, A. J. Williams, P. W. Stephens, J. Tao, Y. Zhu, V. Ksenofontov, F. Casper, C. Felser, and R. J. Cava, Phys. Rev. Lett. **103**, 057002 (2009).

[11] S. H. Baek, D. V. Efremov, J. M. Ok, J. S. Kim, J. van den Brink, and B. Buechner, Nat. Mater. **14**, 210 (2015).

[12] K. Nakayama, Y. Miyata, G. N. Phan, T. Sato, Y. Tanabe, T. Urata, K. Tanigaki, and T. Takahashi, Phys. Rev. Lett. **113**, 237001 (2014).

[13] T. Shimojima, Y. Suzuki, T. Sonobe, A. Nakamura, M. Sakano, J. Omachi, K. Yoshioka, M. Kuwata-Gonokami, K. Ono, H. Kumigashira, A. E. Boehmer, F. Hardy, T. Wolf, C. Meingast, H. V. Loehneysen, H. Ikeda, and K. Ishizaka, Phys. Rev. B **90**, 121111 (R) (2014).

[14] J. Zhao, Q. Huang, C. de la Cruz, S. L. Li, J. W. Lynn, Y. Chen, M. A. Green, G. F. Chen, G. Li, Z. Li, J. L. Luo, N. L. Wang, and P. C. Dai, Nat. Mater. **7**, 953 (2008).

[15] Q. Huang, Y. Qiu, W. Bao, M. A. Green, J. W. Lynn, Y. C. Gasparovic, T. Wu, G. Wu, and X. H. Chen, Phys. Rev. Lett. **101**, 257003 (2008).

[16] M. D. Watson, T. K. Kim, A. A. Haghighirad, N. R. Davies, A. McCollam, A. Narayanan, S. F. Blake, Y. L. Chen, S. Ghannadzadeh, A. J. Schofield, M. Hoesch, C. Meingast, T. Wolf, and A. I. Coldea, Phys. Rev. B **91**, 155106 (2015).

[17] M. A. Tanatar, A. E. Bohmer, E. I. Timmons, M. Schutt, G. Drachuck, V. Taufour, K. Kothapalli, A. Kreyssig, S. L. Bud'ko, P. C. Canfield, R. M. Fernandes, and R. Prozorov, Phys. Rev. Lett. **117**, 127001 (2016).

[18] D. N. Yuan, J. Yuan, Y. L. Huang, S. L. Ni, Z. P. Feng, H. X. Zhou, Y. Y. Mao, K. Jin, G. M. Zhang, X. L. Dong, F. Zhou, and Z. X. Zhao, Phys. Rev. B **94**, 060506 (R) (2016).

[19] T. Hashimoto, Y. Ota, H. Q. Yamamoto, Y. Suzuki, T. Shimojima, S. Watanabe, C. Chen, S. Kasahara, Y. Matsuda, T. Shibauchi, K. Okazaki, and S. Shin, Nat. Commun. **9**, 282 (2018).

[20] D. F. Liu, C. Li, J. W. Huang, B. Lei, L. Wang, X. X. Wu, B. Shen, Q. Gao, Y. X. Zhang, X. Liu, Y. Hu, Y. Xu, A. J. Liang, J. Liu, P. Ai, L. Zhao, S. L. He, L. Yu, G. D. Liu, Y. Y. Mao, X. L. Dong, X. W. Jia, F. F. Zhang, S. J. Zhang, F. Yang, Z. M. Wang, Q. J. Peng, Y. G. Shi, J. P. Hu, T. Xiang, X. H.

Chen, Z. Y. Xu, C. T. Chen, and X. J. Zhou, Phys. Rev. X **8**, 031033 (2018).

[21] J. K. Glasbrenner, I. I. Mazin, H. O. Jeschke, P. J. Hirschfeld, R. M. Fernandes, and R. Valenti, Nat. Phys. **11**, 953 (2015).

[22] S. Onari, Y. Yamakawa, and H. Kontani, Phys. Rev. Lett. **116**, 227001 (2016).

[23] A. V. Chubukov, M. Khodas, and R. M. Fernandes, Phys. Rev. X **6**, 041045 (2016).

[24] M. D. Watson, T. Yamashita, S. Kasahara, W. Knafo, M. Nardone, J. Beard, F. Hardy, A. McCollam, A. Narayanan, S. F. Blake, T. Wolf, A. A. Haghighirad, C. Meingast, A. J. Schofield, H. v. Loehneysen, Y. Matsuda, A. I. Coldea, and T. Shibauchi, Phys. Rev. Lett. **115**, 027006 (2015).

[25] K. K. Huynh, Y. Tanabe, T. Urata, H. Oguro, S. Heguri, K. Watanabe, and K. Tanigaki, Phys. Rev. B **90**, 144516 (2014).

[26] H. Sun, D. N. Woodruff, S. J. Cassidy, G. M. Allcroft, S. J. Sedlmaier, A. L. Thompson, P. A. Bingham, S. D. Forder, S. Cartenet, N. Mary, S. Ramos, F. R. Foronda, B. H. Williams, X. Li, S. J. Blundell, and S. J. Clarke, Inorg. Chem. **54**, 1958 (2015).

[27] The Supplemental Material (http://---) includes the Methods, supplemental figures (XRD at 300 K, Williamson-Hall plot, $\rho_{xy}(H)$ of H7.2 and C8.6, $\rho(T)$ of pressurized H7.2, correlation between $x$ and $c$), and supplemental tables (results of structural refinements and Williamson-Hall analysis).

[28] T. M. McQueen, Q. Huang, V. Ksenofontov, C. Felser, Q. Xu, H. Zandbergen, Y. S. Hor, J. Allred, A. J. Williams, D. Qu, J. Checkelsky, N. P. Ong, and R. J. Cava, Phys. Rev. B **79**, 014522 (2009).

[29] E. Pomjakushina, K. Conder, V. Pomjakushin, M. Bendele, and R. Khasanov, Phys. Rev. B **80**, 024517 (2009).

[30] A. E. Boehmer, F. Hardy, F. Eilers, D. Ernst, P. Adelmann, P. Schweiss, T. Wolf, and C. Meingast, Phys. Rev. B **87**, 180505 (R) (2013).

[31] C. Koz, M. Schmidt, H. Borrmann, U. Burkhardt, S. Roessler, W. Carrillo-Cabrera, W. Schnelle, U. Schwarz, and Y. Grin, Z. Anorg. Allg. Chem. **640**, 1600 (2014).

[32] Z. P. Feng, J. Yuan, X. X. W. J. Li, W. Hu, B. Shen, M. Y. Qin, L. Zhao, B. Y. Zhu, V. Stanev, M. Liu, G. M. Zhang, X. L. Dong, F. Zhou, X. J. Zhou, J. P. Hu, I. Takeuchi, Z. X. Zhao, and K. Jin, arXiv1807.01273.

[33] S. L. Ni, W. Hu, P. P. Shen, Z. X. Wei, S. B. Liu, D. Li, J. Yuan, L. Yu, K. Jin, F. Zhou, X. L. Dong, and Z. X. Zhao, Chin. Phys. B **28**, 127401 (R) (2019).

[34] F. Hardy, A. E. Boehmer, D. Aoki, P. Burger, T. Wolf, P. Schweiss, R. Heid, P. Adelmann, Y. X. Yao, G. Kotliar, J. Schmalian, and C. Meingast, Phys. Rev. Lett. **111**, 027002 (2013).

[35] Y. P. Wu, D. Zhao, A. F. Wang, N. Z. Wang, Z. J. Xiang, X. G. Luo, T. Wu, and X. H. Chen, Phys. Rev. Lett. **116**, 147001 (2016).

[36] M. Q. He, L. R. Wang, F. Hardy, L. P. Xu, T. Wolf, P. Adelmann, and C. Meingast, Phys. Rev. B **97**, 104107 (2018).

[37] G. M. Zhang, Y. H. Su, Z. Y. Lu, Z. Y. Weng, D. H. Lee, and T. Xiang, Europhys. Lett. **86**, 37006 (2009).

[38] J. P. Sun, K. Matsuura, G. Z. Ye, Y. Mizukami, M. Shimozawa, K. Matsubayashi, M. Yamashita, T. Watashige, S. Kasahara, Y. Matsuda, J. Q. Yan, B. C. Sales, Y. Uwatoko, J. G. Cheng, and T. Shibauchi, Nat. Commun. **7**, 12146 (2016).

[39] K. Kothapalli, A. E. Bohmer, W. T. Jayasekara, B. G. Ueland, P. Das, A. Sapkota, V. Taufour, Y. Xiao, E. Alp, S. L. Bud'ko, P. C. Canfield, A. Kreyssig, and A. I. Goldman, Nat. Commun **7**, 12728 (2016).

[40] J. P. Sun, G. Z. Ye, P. Shahi, J. Q. Yan, K. Matsuura, H. Kontani, G. M. Zhang, Q. Zhou, B. C.

Sales, T. Shibauchi, Y. Uwatoko, D. J. Singh, and J. G. Cheng, Phys. Rev. Lett. **118**, 147004 (2017).

[41] E. Dagotto, Rev. Mod. Phys. **66**, 763 (1994).

[42] H. Takagi, S. Uchida, and Y. Tokura, Phys. Rev. Lett. **62**, 1197 (1989).

[43] T. K. Chen, C. C. Chang, H. H. Chang, A. H. Fang, C. H. Wang, W. H. Chao, C. M. Tseng, Y. C. Lee, Y. R. Wu, M. H. Wen, H. Y. Tang, F. R. Chen, M. J. Wang, M. K. Wu, and D. Van Dyck, Proc. Natl. Acad. Sci. U. S. A. **111**, 63 (2014).

[44] M. Kawaminami and A. Okazaki, J. Phys. Soc. Jpn. **29**, 649 (1970).

[45] K. Haule and G. Kotliar, New J. Phys. **11**, 025021 (2009).

**Additional information**

Supplemental information is available in the online version of the paper.

**Competing financial interests**

The authors declare no competing financial interests.

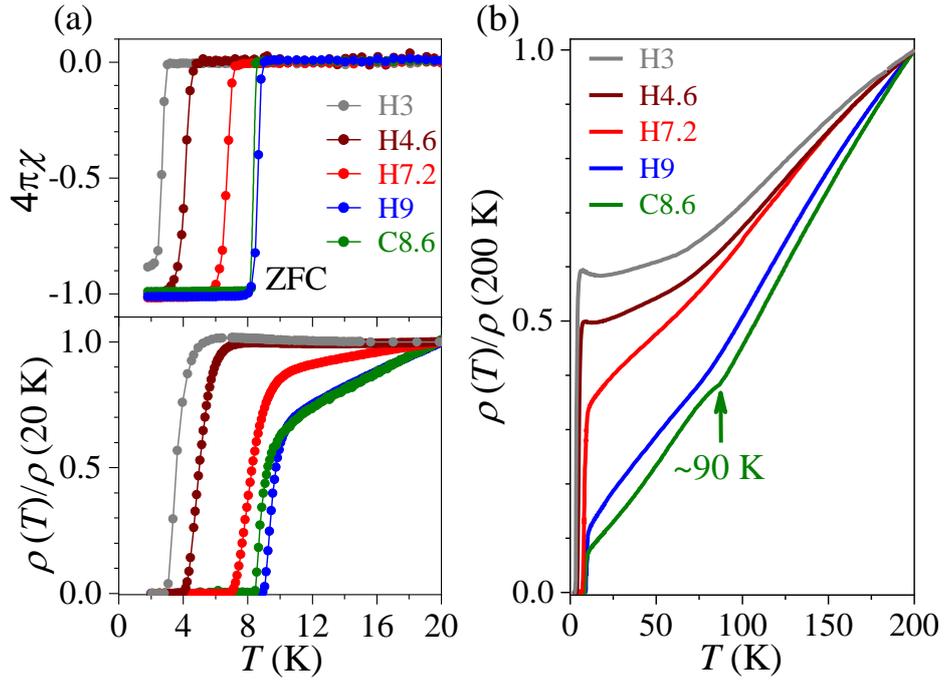

FIG. 1. (a) Temperature dependences of magnetic susceptibility (the upper panel) and scaled in-plane electrical resistivity $\rho(T)$ up to 20 K (the lower panel) for representative HID and CVT samples. All the samples show the sharp bulk superconducting transitions with ~100 % shielding signals. The magnetic data were measured in zero-field-cooling (ZFC) mode in a 1 Oe field parallel to $c$ axis, and corrected for demagnetization factor. (b) The scaled $\rho(T)$ curves up to 200 K.

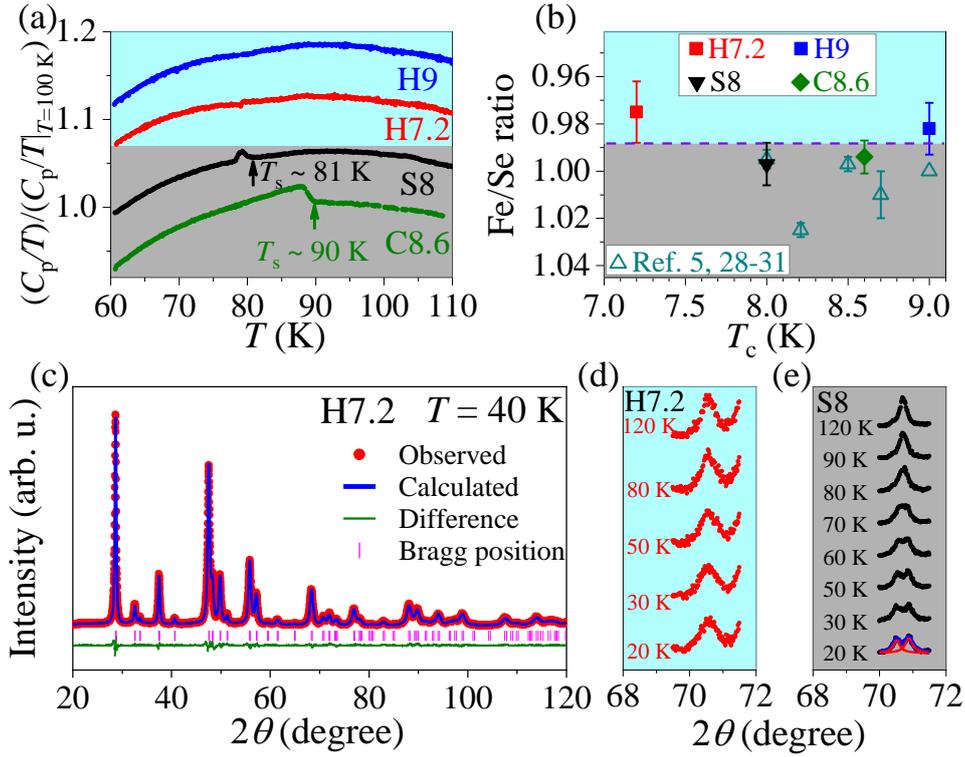

FIG. 2. (a) Temperature dependences of specific heat for HID (H9, H7.2) and CVT (C8.6) single crystals, as well as polycrystalline sample S8. The curves are offset for clarity. Different background colors are used for the samples having (gray) or not having (blue) the structural transition at $T_s$. (b) Plot of Fe/Se ratio vs. $T_c$ among the present HID, CVT, and polycrystalline samples, as well as the samples of previous work [5,28-31]. The background color codes are the same as in (a). (c) Powder XRD pattern at 40 K and corresponding Rietveld refined profile (see Table S2 for the refinement results) for HID H7.2. (d), (e) Comparison between the single tetragonal (220) peak down to 20 K by the powder XRD for HID H7.2 and the peak splitting into the orthorhombic (400)/(040) doublet for polycrystalline sample S8.

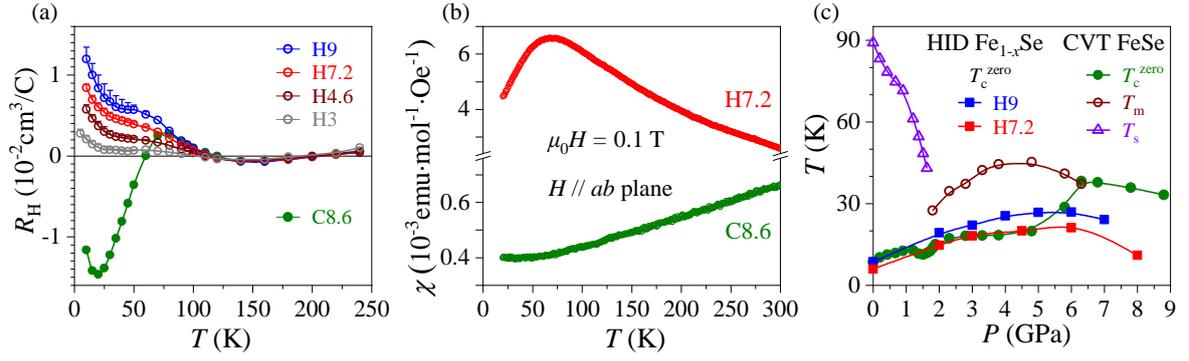

FIG. 3. (a) Temperature dependences of Hall coefficient $R_H$ for HID Fe$_{1-x}$Se and CVT FeSe single crystals. Not all the data are shown for clarity. Here Hall coefficient is defined as the field derivative of Hall resistivity $\rho_{xy}(H)$, $R_H = d\rho_{xy}(H)/dH$, at the zero-field limit. The small error bars of the HID samples indicate a slight nonlinearity in their low-T $\rho_{xy}(H)$ curves. (b) Temperature dependences of normal-state magnetic susceptibility up to room temperature for H7.2 and C8.6 single crystals. (c) T-P phase diagrams of HID (H7.2, H9) and CVT [38,40] single crystals. The data of HID samples are obtained by measuring the in-plane resistivity $\rho(T)$ under high hydrostatic pressures [27]. Here the SC transition temperature is defined as the onset temperature of zero resistivity, $T_c^{zero}$.

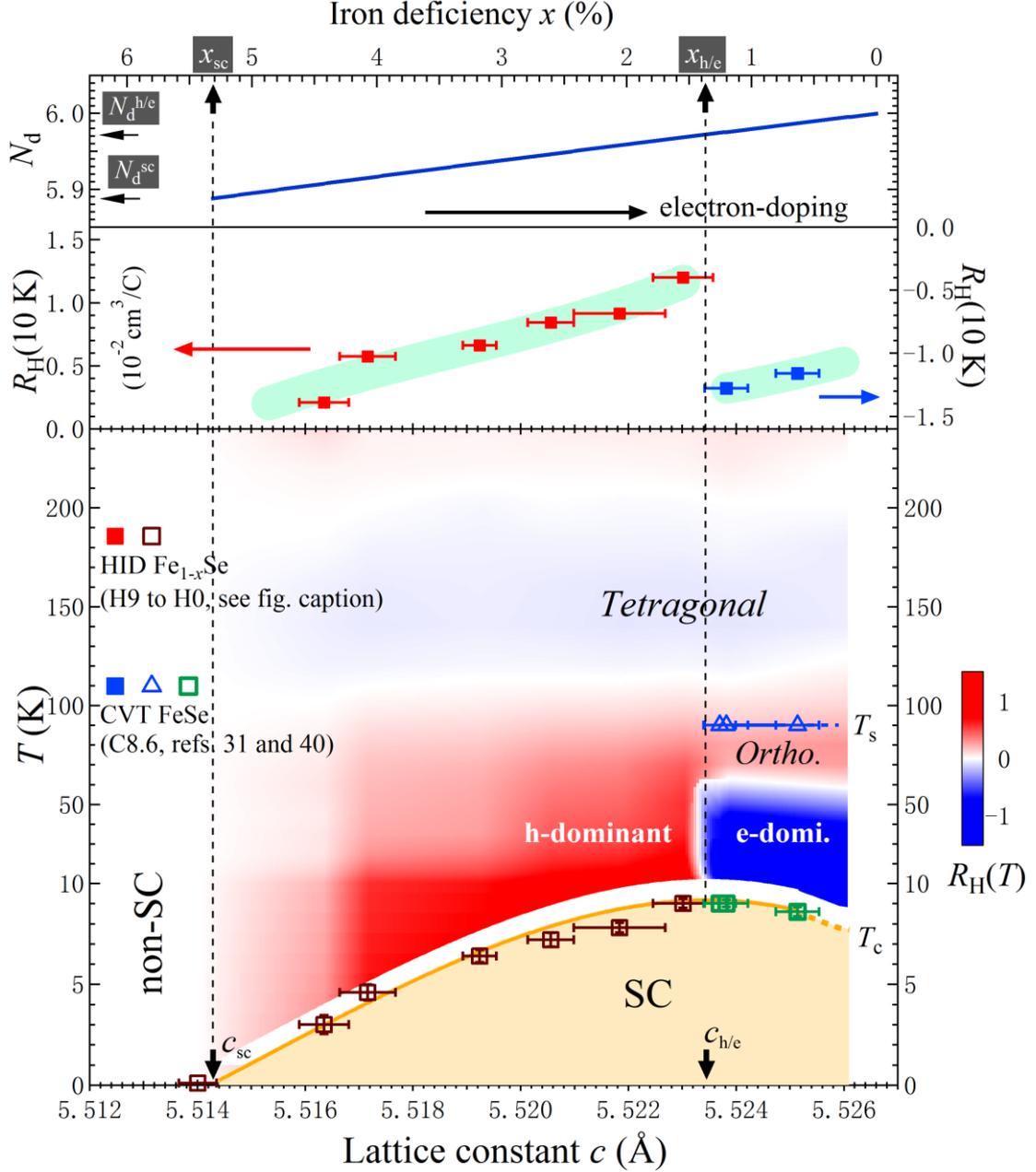

Fig. 4. **Lower panel:** Phase diagram of FeSe system, plotted by temperature $T$ vs. lattice parameter $c$ (at 300 K) and superimposed by a contour plot of Hall coefficient $R_H(T)$ in the normal state. The SC transition width is indicated by the vertical error bars of the $T_c$ data. The solid/dashed orange curve is a guide to the eye. The Fe-deficiency $x$ (the upper horizontal axis) correlates with the lattice parameter $c$, as inferred from Fig. S5. The phase diagram is constructed with the data of seven HID $Fe_{1-x}Se$ (H0, H3, H4.6, H6.4, H7.2, H7.8, H9) and three CVT FeSe (C8.6, plus the other two with $T_c$ = 9 K from refs. 31 and 40) single crystals. HID sample H0 shows no superconductivity down to 1.8 K. **Middle panel:** The $c$ dependence of Hall coefficient $R_H(T)$ at low $T$ = 10 K just above $T_c$. The $R_H$(10 K) values are positive for $x \gtrsim x_{h/e}$ ($N_d \lesssim N_d^{h/e}$) and negative for $x \lesssim x_{h/e}$ ($N_d \gtrsim N_d^{h/e}$). **Upper panel:** The $d$-electron number per iron ($N_d$) in $Fe_{1-x}Se$ as a function of Fe-deficiency $x$, i.e. $N_d$ = 6 - $p$ = (6-8$x$)/(1-$x$), in the range of $x_{sc} \geq x \geq 0$.